\documentstyle[12pt]{article}

\catcode`\@=11
\long\def\@makefntext#1{
\protect\noindent \hbox to 3.2pt {\hskip-.9pt
$^{{\ninerm\@thefnmark}}$\hfil}#1\hfill}                %CAN BE USED

\def\@makefnmark{\hbox to 0pt{$^{\@thefnmark}$\hss}}  %ORIGINAL

\def\ps@myheadings{\let\@mkboth\@gobbletwo
\def\@oddhead{\hbox{}
\rightmark\hfil\ninerm\thepage}
\def\@oddfoot{}\def\@evenhead{\ninerm\thepage\hfil
\leftmark\hbox{}}\def\@evenfoot{}
\def\sectionmark{}\def\subsectionmark{}}

\renewcommand{\thefootnote}{\fnsymbol{footnote}}

\renewcommand{\subsubsection}[1]
{\vspace*{0.6cm}\addtocounter{subsubsectionc}{1}
        \noindent
{\normalsize\rm\thesectionc.\thesubsectionc.\thesubsubsectionc.
        #1}\par\vspace*{0.6cm}}

\newcounter{appendixc}
\newcounter{subappendixc}[appendixc]
\newcounter{subsubappendixc}[subappendixc]

\renewcommand{\appendix}[1] {\vspace*{0.6cm}
        \refstepcounter{appendixc}
        \setcounter{figure}{0}
        \setcounter{table}{0}
        \setcounter{equation}{0}
        \renewcommand{\thefigure}{\Alph{appendixc}.\arabic{figure}}
        \renewcommand{\thetable}{\Alph{appendixc}.\arabic{table}}
        \renewcommand{\theappendixc}{\Alph{appendixc}}
        \renewcommand{\theequation}{\Alph{appendixc}.\arabic{equation}}
        \noindent{\bf Appendix \theappendixc #1}\par\vspace*{0.4cm}}

\def\abstracts#1{{

\centering{\begin{minipage}{12.2truecm}\footnotesize\baselineskip=12pt\noindent
        \centerline{\footnotesize ABSTRACT}\vspace*{0.3cm}
        \parindent=0pt #1
        \end{minipage}}\par}}

\renewenvironment{thebibliography}[1]
        {\begin{list}{\arabic{enumi}.}
        {\usecounter{enumi}\setlength{\parsep}{0pt}
\setlength{\leftmargin 1.25cm}{\rightmargin 0pt}
         \setlength{\itemsep}{0pt} \settowidth
        {\labelwidth}{#1.}\sloppy}}{\end{list}}

\newcounter{itemlistc}
\newcounter{romanlistc}
\newcounter{alphlistc}
\newcounter{arabiclistc}

\newcommand{\fcaption}[1]{
        \refstepcounter{figure}
        \setbox\@tempboxa = \hbox{\footnotesize Fig.~\thefigure. #1}
        \ifdim \wd\@tempboxa > 6in
           {\begin{center}
        \parbox{6in}{\footnotesize\baselineskip=12pt Fig.~\thefigure. #1}
            \end{center}}
        \else
             {\begin{center}
             {\footnotesize Fig.~\thefigure. #1}
              \end{center}}
        \fi}

\newcommand{\tcaption}[1]{
        \refstepcounter{table}
        \setbox\@tempboxa = \hbox{\footnotesize Table~\thetable. #1}
        \ifdim \wd\@tempboxa > 6in
           {\begin{center}
        \parbox{6in}{\footnotesize\baselineskip=12pt Table~\thetable. #1}
            \end{center}}
        \else
             {\begin{center}
             {\footnotesize Table~\thetable. #1}
              \end{center}}
        \fi}

\def\@citex[#1]#2{\if@filesw\immediate\write\@auxout
        {\string\citation{#2}}\fi
\def\@citea{}\@cite{\@for\@citeb:=#2\do
        {\@citea\def\@citea{,}\@ifundefined
        {b@\@citeb}{{\bf ?}\@warning
        {Citation `\@citeb' on page \thepage \space undefined}}
        {\csname b@\@citeb\endcsname}}}{#1}}

\newif\if@cghi
\def\cite{\@cghitrue\@ifnextchar [{\@tempswatrue
        \@citex}{\@tempswafalse\@citex[]}}
\def\citelow{\@cghifalse\@ifnextchar [{\@tempswatrue
        \@citex}{\@tempswafalse\@citex[]}}
\def\@cite#1#2{{$\null^{#1}$\if@tempswa\typeout
        {IJCGA warning: optional citation argument
        ignored: `#2'} \fi}}

 1
 1
 1

\font\ninerm=cmr9

\begin{document}

\centerline{\normalsize\bf RENORMALONS AND PERTURBATIVE FIXED POINTS\footnote
{Invited talk at the International Symposium on Heavy Flavor and Electroweak
Theory,17-19 August 1995,Beijing,China.}}
\baselineskip=22pt

\vspace{0.6cm}
\centerline{\footnotesize Georges GRUNBERG}

\baselineskip=13pt
\centerline{\footnotesize\it Centre de Physique Th\'eorique de l'Ecole
Polytechnique\footnote{CNRS UPRA 0014}}
\baselineskip=12pt
\centerline{\footnotesize\it 91128 Palaiseau Cedex - France}
\vspace{0.6cm}
\centerline{\footnotesize E-mail: Grunberg@orphee.polytechnique.fr}

\vspace{0.9cm}
\abstracts{The connection between renormalons and power corrections is
investigated
for the typical infrared renormalon integral assuming the effective
coupling constant has an infrared fixed
point of an entirely perturbative origin. It is shown the full answer
differs from the Borel sum by a power correction.}
\newpage
%\vspace*{0.6cm}
\normalsize\baselineskip=15pt
\setcounter{footnote}{0}
\renewcommand{\thefootnote}{\alph{footnote}}
%\vspace{0.9cm}
It is generally believed that the large order behavior associated to
infrared (IR) renormalons,
which makes QCD perturbation theory "non Borel summable", reflects an
inconsistency related to
the Landau ghost. In this note, I examine how the renormalon problem is
resolved when
the coupling constant is free of Landau singularity and approaches a non
trivial IR
fixed point at small momenta. I assume the fixed point arises entirely
within a perturbative
framework, through higher order perturbative corrections. I will show that
in this case, IR
renormalons are also present, but the exact result differs from the Borel
sum by a (complex)
power correction, which removes all inconsistencies.

Let us first review the standard argument[1,2] for renormalons. Consider
the typical IR renormalon
integral :

\begin{eqnarray}
R(\alpha) = \int_{0}^{Q^{2}} n \frac{dk^{2}}{k^{2}}
\left(\frac{k^{2}}{Q^{2}}\right)^{n}
\alpha_{eff} (k/Q,\alpha) %(1)
\end{eqnarray}
where $\alpha$ is the coupling at scale $Q$ in some arbitrary renormalization
scheme, and $\alpha_{eff}(k)$ a renormalization group (RG) invariant
effective coupling
(I assume $n > 0$, so that the integral in eq.(1) is IR convergent order by
order
in perturbation theory). Let us compute the perturbative Borel transform
$R(z)$. If
$R(\alpha)$ has the formal power series expansion :

$$
{R_{PT}(\alpha) = \mathop{\sum}\limits_{p=0}^{\infty} r_p\ \alpha^{p+1},}
$$
$R(z)$ is defined  by  :
\begin{equation}
R(z) = \mathop{\sum}\limits_{p=0}^{\infty}\ \frac{r_p}{p!}\ z^p %(2)
\end{equation}
The series eq.(2) are believed to have a finite convergence
radius (at the difference of those for $R(\alpha)$), and therefore
allow for an all order definition of $R(z)$. One can then {\sl define} an
all order "perturbative" resummed $R_{PT}(\alpha)$ by the Borel
representation~:

\begin{eqnarray}
R_{PT}(\alpha) = \int_0^{\infty} dz\ exp \left(- \frac{z}{\alpha}\right)
R(z) %(3)
\end{eqnarray}
It is convenient to express $R(z)$ in term of the Borel transform of
$\alpha_{eff}$ :

\begin{eqnarray}
\alpha_{eff}(k/Q,\alpha) = \int_0^{\infty} dz\ exp \left(-
\frac{z}{\alpha}\right)
\alpha_{eff}(k/Q,z) %(4)
\end{eqnarray}
Inserting eq.(4) into eq.(1), and interchanging the order of integrations,
one indeed
recovers eq.(3) with $R(\alpha) = R_{PT}(\alpha)$ and :

\begin{eqnarray}
R(z) = \int_0^{Q^2} n \frac{dk^2}{k^2} \left(\frac{k^2}{Q^2}\right)^n
\alpha_{eff}(k/Q,z) %(5)
\end{eqnarray}%(5)
IR renormalons arise as an IR divergence [1] of the integral in eq.(5),
resulting from the IR behavior
of $\alpha_{eff} (k/Q,z)$ (which follows solely from the RG invariance of
$\alpha_{eff}$)~:

\begin{eqnarray}
\alpha_{eff} (k/Q,z) &\simeq &\alpha_{eff}(z) exp
\left\lbrack -z \left(\beta_0\ ln \left(\frac {k^2}{Q^2}\right)
- \frac{\beta_1}{\beta_0} ln\ ln
\left(\frac{Q^2}{k^2}\right)\right)\right\rbrack\nonumber\\
 k^2 &\ll &Q^2
\end{eqnarray}%(6)
where $\beta_0$ and $\beta_1$ are (minus) the one and two loop beta
function coefficients.
Using eq.(6) into eq.(5) one indeed finds for $z \rightarrow z_n = n/\beta_0$ :

\begin{eqnarray}
R(z) \simeq \alpha_{eff} (z \simeq z_n) \frac {\Gamma(1+\delta)} {n^{\delta}}
\ \frac{1}{(1 - \frac{z}{z_n})^{1+\delta}}
\end{eqnarray}
with $\delta = \frac{\beta_1}{\beta_0} z_n$, i.e. $R(z)$ displays
a cut singularity [2] at the renormalon position $z = z_n$, which generates
according to eq.(3), since $z_n > 0$, an $O(exp(-z_n/\alpha))$ imaginary part
in $R_{PT}(\alpha)$. In the standard case where $\alpha_{eff}$ has a Landau
ghost
at some scale $\Lambda^2 < Q^2$, this fact causes no surprise, since the
defining
integral eq.(1) itself involves an ambiguous integration over the Landau
singularity. But a paradox arises if $\alpha_{eff}$ has an IR fixed point :
then
$R(\alpha)$ should be perfectly well defined, with no imaginary part
according to
eq.(1), whereas eq.(3), together with eq.(7), still yield an ambiguous,
imaginary
amplitude for $R_{PT}(\alpha)$ ! But eq.(7) is correct in both cases : as
mentionned
above, it follows solely from $RG$ invariance and the representation eq.(5)
(which
is presumably always valid since it is correct order by order in perturbation
expansion in $z$, and the corresponding series are convergent). Thus $IR$
renormalons
{\em are also present in the IR fixed point case}. The only possible
conclusion is that eq.(3)
is not in fact a valid representation of eq.(1), i.e. that $R(\alpha)$ in
eq.(1) does not
coincide with $R_{PT}(\alpha)$ in eq.(3) in the latter case. This is
despite the fact I
assumed eq.(4) does instead correctly represents (at least for large enough
$k$) $\alpha_{eff}(k)$
(which is also assumed to have no renormalons), which can thus be
determined in principle
from "all order" perturbation theory, and qualifies the present framework
as being
"perturbative".

Indeed, there is a loophole in the previous derivation that $R(\alpha) =
R_{PT}(\alpha)$,
which is actually already present when there is a Landau singularity.
Assume e.g. $\alpha_{eff}$
is the one loop coupling :

\begin{eqnarray}
\alpha_{eff}(k/Q,\alpha) & = & {\displaystyle\frac {\alpha}{1 +
\alpha\beta_0\ ln \left(\frac{k^2}{Q^2}\right)}}
\nonumber\\
& \equiv & \frac{1}{\beta_0\ ln \left(\frac{k^2}{\Lambda^2}\right)}\nonumber
\end{eqnarray}

Then [1] : $\alpha_{eff}(k/Q,z) = exp \lbrack -z\ \beta_0\ ln
\left(\frac{k^2}{Q^2}\right)\rbrack$ and eq.(5) yields :

$$
R(z) = \int_0^{Q^{2}} n \frac{dk^2}{k^2} \left(\frac{k^2}{Q^2}\right)^n
exp \left\lbrack -z\ \beta_0\ ln\ \left(\frac{k^2}{Q^2}\right)\right\rbrack
= \frac{1}{1-\frac{z}{z_n}}
$$
The problem with the previous argument is that eq.(4) itself is a valid
representation
of $\alpha_{eff}(k)$ {\sl only if $k$ is not too small}. Indeed in the
present example the integral in eq.(4) converges at $z = \infty$
only if $\frac{1}{\alpha} + \beta_0\ ln \left(\frac{k^2}{Q^2}\right) > 0$,
i.e. for $k^2 > \Lambda^2$. For $k^2 < \Lambda^2$, $\alpha_{eff}(k)$
has to be represented instead by a Borel integral over the {\sl negative}
$z$ axis :

\begin{eqnarray}
\alpha_{eff}(k/Q,\alpha) = - \int_{-\infty}^0 dz\ exp\
\left(-\frac{z}{\alpha}\right)
\alpha_{eff} (k/Q,z)
\end{eqnarray}
Similar remarks apply in the general case, where eq.(6) implies, provided
$\alpha_{eff}(z)$
decreases no faster then $exp(-cz)$ at large $z$,\ $\alpha_{eff}(k)$ will have
a Landau singularity at some scale $k^2 = \Lambda^2$ (see however the fixed
point
case), below which the Borel representation eq.(4) will break down, and eq.(8)
should be used instead.

These observations suggest that the correct procedure in the Landau ghost
case is to first
take $\alpha < 0$ (and slightly complex if $\beta_1 \not= 0$), which
implies $Q^2 < \Lambda^2$,
so that in the whole integration range in eq.(1) one has $k^2 < Q^2 <
\Lambda^2$, and the
representation eq.(8) is valid {\sl throughout} the range (this means
choosing $\alpha$ in the domain
of attraction of the {\sl trivial} $IR$ fixed point). Manipulations
analoguous to those performed
above are now justified, and yield :

\begin{eqnarray}
R(\alpha) = - \int_{-\infty}^0 dz\ exp\ \left(- \frac {z}{\alpha}\right) R(z)
\end{eqnarray}%(9)
Finally, $R(\alpha)$ for $\alpha > 0$ (where $Q^2 > \Lambda^2$) is defined
as the analytic
continuation of eq.(9), which yields the standard Borel representation
eq.(3) with $R = R_{PT}$.

In the case where $\alpha_{eff}(k)$ has an IR fixed point, a similar
problem arises if it happens
again that the representation eq.(4) is not valid below some scale $k =
k_{min}$, even if $k_{min}$
does not correspond to a Landau singularity this time. This is possible if,
as in the Landau ghost case,
$\alpha_{eff}(z)$ does not decrease too fast at large $z$ (the latter
assumption seems necessary,
because a too fast decrease of $\alpha_{eff}(z)$ (e.g. [1] if
$\alpha_{eff}(z) = exp(-cz^2)$), although
insuring the validity of eq.(4) for all k's, usually yields an
$\alpha_{eff}(k)$ which blows up too fast
as $k \rightarrow 0$, making the original integral eq.(1) IR divergent).
Furthermore, {\sl and this is the crucial
difference with the Landau ghost case}, for $k < k_{min}$\ ,
$\alpha_{eff}(k)$, which remains positive,
will not admit a Borel representation over the $z < 0$ axis either. (Note
that for $k \rightarrow 0$,
$\alpha_{eff}(k)$ is always in the strong coupling region, since it
approaches a non-trivial fixed point,
at the difference of the Landau ghost case, where for $k \rightarrow 0$ one
enters an (IR) asymptotically
free region below the Landau singularity, since no other fixed point is
available). The above derivation that
$R = R_{PT}$ can thus not be extended to the IR fixed point case, as
expected. Let us now give a general
argument that these two functions actually differ by a power correction.
Splitting the integration range
in eq.(1) at $k = k_{min}$, one has :

\begin{eqnarray}
R(\alpha) & = & \int_0^{k_{min}^2} n\ \frac{dk^2}{k^2}
\left(\frac{k^2}{Q^2}\right)^n
\alpha_{eff} \left(\frac{k}{Q},\alpha\right) + \int_{k_{min}^2}^{Q^2} n\
\frac{dk^2}{k^2}
\left(\frac{k^2}{Q^2}\right)^n \alpha_{eff}
\left(\frac{k}{Q},\alpha\right)\nonumber\\
& \equiv &R_- + R_+ \nonumber
\end{eqnarray}
Eq.(4) cannot be used inside the low momentum integral $R_-$, which can
however be parametrized
as a power correction since :

\begin{eqnarray}
R_- & = & \left(\frac{k_{min}^2}{Q^2}\right)^n \int_0^{k_{min}^2} n\
\frac{dk^2}{k^2}
\left(\frac{k^2}{k_{min}^2}\right)^n \alpha_{eff} \left(\frac{k}{k_{min}} ,
\alpha_{min}\right)\nonumber\\
& \equiv & \left(\frac{k_{min}^2}{Q^2}\right)^n R(\alpha_{min})\nonumber
\end{eqnarray}
where $\alpha_{min} = \alpha(Q = k_{min})$, and RG invariance has been used
in the first step.
On the other hand, using eq.(4) in $R_+$, one gets :

$$
R_+ = \int_0^{\infty} dz\ exp \left(-\frac{z}{\alpha}\right)
\int_{k_{min}^2}^{Q^2} n\ \frac{dk^2}{k^2} \left(\frac{k^2}{Q^2}\right)^n
\alpha_{eff} \left(\frac{k}{Q},z\right)
$$
To go further, it is necessary to know the k-dependence of
$\alpha_{eff}(k/Q,z)$. This can be
done easily in the special case where $\beta_1 = 0$, if one chooses
$\alpha(Q)$ to be the one
loop coupling. Then one can show[1] that : $\alpha_{eff}(k/Q,z) =
\alpha_{eff}(z) exp\lbrack -z \beta_0\ ln (k^2/Q^2)\rbrack$,
and one gets :

$$
\int_{k_{min}^2}^{Q^2} n\ \frac{dk^2}{k^2} \left(\frac{k^2}{Q^2}\right)^n
\alpha_{eff} \left(\frac{k}{Q},z\right) = \alpha_{eff}
\frac{1}{1-\frac{z}{z_n}}
\left\lbrack 1 - \left(\frac{k_{min}^2}{Q^2}\right)^{z_n \beta_0
(1-z/z_n)}\right\rbrack
$$
Hence :

\begin{eqnarray}
R_+ & = &\int_0^{\infty} dz\ exp \left(-\frac{z}{\alpha}\right)
\alpha_{eff}(z) \frac{1}{1 - \frac{z}{z_n}}
- \left(\frac{k_{min}^2}{Q^2}\right)^n \int_0^{\infty} dz\ exp \left(-
\frac{z}{\alpha_{min}} \right)
\alpha_{eff}(z) \frac {1}{1 - \frac{z}{z_n}}\nonumber\\
& \equiv & R_{PT}(\alpha) - \left(\frac{k_{min}^2}{Q^2}\right)^n
R_{PT}(\alpha_{min})\nonumber
\end{eqnarray}
One therefore ends up with the result :

\begin{eqnarray}
R(\alpha) = R_{PT}(\alpha) + \left(\frac{k_{min}^2}{Q^2}\right)^n \lbrack
R(\alpha_{min}) - R_{PT}(\alpha_{min})\rbrack
\end{eqnarray}
where the coefficient of the power correction is given by the discrepancy
between the exact
amplitude and its Borel representation. Eq.(10) is equivalent to the
statement that
$R(\alpha) = R_{PT}(\alpha) + const/(Q^2)^n$, and is also correct when
$\beta_1 \not= 0$ (see below).
Note it is not possible the power correction vanishes (as it does in the
Landau ghost case)
since $R(\alpha_{min})$ is real, whereas $R_{PT}(\alpha_{min})$ is complex
due to the effect of the renormalon (the power correction must be
complex to cancell the imaginary part of $R_{PT}(\alpha))$.

I now illustrate the previous discussion with the example of the 2 loop
coupling :

\begin{eqnarray}
\frac{d\alpha_{eff}}{d ln k^2} = - \beta_0 (\alpha_{eff})^2 - \beta_1
(\alpha_{eff})^3
\end{eqnarray}
I shall consider both the standard case $\beta_1/\beta_0 > 0$ where there
is a Landau singularity,
and the case $\beta_1/\beta_0 < 0$ where $\alpha_{eff}(k)$ has an IR fixed
point at $\alpha_{IR} =
 -\beta_0/\beta_1$ (which actually occurs in QCD for a large enough number
of flavors). Remarkably,
$R(z)$ can be computed {\sl exactly} with a straightforward change of
variable, adapted from a similar one suggested in ref.[3].
Defining the Borel variable by :

\begin{eqnarray}
\frac{z}{z_n} = \frac{1 - \frac{\alpha}{\alpha_{eff}(k)}}{1 +
\frac{\beta_1}{\beta_0} \alpha}
\end{eqnarray}
with $\alpha \equiv \alpha_{eff} (k = Q)$, and using the solution of eq.(11) :

\begin{eqnarray}
ln \left(\frac{k^2}{\Lambda^2}\right) = \frac {1}{\beta_0 \alpha_{eff}} -
\frac{\beta_1}{\beta_0^2}
\ ln \left(\frac{1}{\alpha_{eff}} + \frac{\beta_1}{\beta_0}\right) +
\frac{\beta_1}{\beta_0^2}
\end{eqnarray}%(13)
(together with the similar relation at $k = Q$) one obtains :

$$
n \frac{dk^2}{k^2} \left(\frac{k^2}{Q^2}\right)^n \alpha_{eff}(k) = - dz
\frac{exp (-\frac{z}{\alpha}
- \frac{\beta_1}{\beta_0} z)}{\left(1 - \frac{z}{z_n}\right)^{1 + \delta}}
$$
which suggests the looked for Borel transform is :

\begin{eqnarray}
R(z) = \frac{exp \left(- \frac{\beta_1}{\beta_0} z\right)}{\left(1 -
\frac{z}{z_n}\right)^{1+\delta}}\ .
\end{eqnarray}
To complete the proof, it remains to determine the new integration bounds.
\vskip 1truecm
1) Assume first $\beta_1 / \beta_0 > 0$ : then $\alpha_{eff}$ has a Landau
singularity
at the scale ${\bar\Lambda}^2 = \Lambda^2 exp \left(- \frac{\beta_1}{\beta_0^2}
ln \left(\frac{\beta_1}{\beta_0}\right) + \frac{\beta_1}{\beta_0^2}\right)$.
I assume $Q^2 > {\bar\Lambda^2}$, so that $\alpha > 0$. For $k^2 = Q^2$,
$\alpha_{eff} = \alpha$,
and $z = 0$ ; at $k^2 = {\bar\Lambda^2}$ , $\alpha_{eff} = \infty$, and $z
= z_L \equiv
\frac{z_n}{1 + \frac{\beta_1}{\beta_0} \alpha} < z_n$ ; decreasing $k^2$
below ${\bar\Lambda}^2$,
$z$ becomes complex ; finally, for $k^2 \rightarrow 0$, $\alpha_{eff}
\rightarrow 0^-$, and
$z \rightarrow +\infty$. $R(\alpha)$ in eq.(1) then takes the form of a
Borel integral along a
path in the complex $z$-plane, which divides into two complex conjuguate
branches at $z = z_L$,
and can be deformed to above or below the positive real axis to yield the
standard Borel representation
eq.(3) with $R(z)$ as in eq.(14).
\vskip 1truecm
2) On the other hand, if $\beta_1/\beta_0 < 0$, the situation is actually
simpler : as $k^2$ decreases
from $Q^2$ to $0$, $\alpha_{eff}$ increases from $\alpha$ to the IR fixed
point $\alpha_{IR}$, and $z$
increases from $0$ to $z_n$ through real values so that eq.(1) becomes :

\begin{eqnarray}
R(\alpha) = \int_0^{z_n} dz\ exp\ \left(- \frac{z}{\alpha}\right)
\frac{exp\left(-\frac{\beta_1}{\beta_0}\ z\right)}{\left(1 -
\frac{z}{z_n}\right)^{1+\delta}}
\end{eqnarray}%(15)
(the renormalon singularity is integrable at $z = z_n$, since $\delta < 0$
now). We therefore
check that $R(\alpha)$ is {\sl not} given by the Borel sum $R_{PT}(\alpha)$
in the IR fixed point case.
Rather, it is given by $R_{PT}(\alpha)$ minus an {\sl exponentially small}
$0(exp(-z_n/\alpha))$ correction~:

\begin{eqnarray}
R(\alpha) & = & \int_0^{\infty} dz\ exp\ \left(- \frac{z}{\alpha}\right)
\frac{exp\left(-\frac{\beta_1}{\beta_0}\ z\right)}{\left(1 -
\frac{z}{z_n}\right)^{1+\delta}}
- \int_{z_n}^{\infty} dz\ exp\ \left(- \frac{z}{\alpha}\right)
\frac{exp\left(-\frac{\beta_1}{\beta_0}\ z\right)}{\left(1 -
\frac{z}{z_n}\right)^{1+\delta}}\nonumber\\
& \equiv & R_{PT} + R_{NP}%(16)
\end{eqnarray}
The latter is just a (complex) power correction, in accordance with the
general expectation.
Indeed, eq.(15)-(16) simplify by performing the change of coupling : $1/a =
1/\alpha + \beta_1/\beta_0$,
and one gets in particular :

\begin{eqnarray}
R_{NP} = - \int_{z_n}^{\infty} dz\ exp\ \left(-\frac{z}{a} \right)
\frac{1}{\left(1 - \frac{z}{z_n}\right)^{1+\delta}}
& = & - \tilde{C}\ exp\ (-z_n/a)(-1/a)^{\delta}\nonumber\\
& = & - \tilde{C}(-1)^{\delta}\ (\Lambda^2/Q^2)^n
\end{eqnarray}%(17)
with $\tilde{C} = \frac{\beta_0}{\beta_1} \Gamma(1-\delta)(z_n)^{\delta}$,
and eq.(13) at $k = Q$
was used in the last step (in this example, the subtracted term $-R_{NP}$
corresponds exactly
to the "minimal" prescription of ref.[4] to "regularize" IR renormalons).
The same method can deal with the case $\alpha < 0$, where one is in the
domain of attraction
of the {\sl trivial} IR fixed point (provided the condition $1 +
\frac{\beta_1}{\beta_0} \alpha > 0$
is also satisfied if $\beta_1/\beta_0 < 0$). Then $\alpha_{eff}(k)$
monotonously increases from $\alpha$
to $0^-$ as $k^2$ decreases from $Q^2$ to 0, hence $z < 0$ and decreases
from 0 to $-\infty$. Thus :

\begin{eqnarray}
R(\alpha) = \int_{-\infty}^0 dz\ exp\ \left(-\frac{z}{\alpha}\right)
\frac{exp\left(-\frac{\beta_1}{\beta_0} z\right)}
{\left(1 - \frac{z}{z_n}\right)^{1+\delta}}
\end{eqnarray}%(18)
i.e. a Borel integral over the {\sl negative} z axis, in agreement with
eq.(9) (note that eq.(15) is
{\sl not} the analytic continuation of eq.(18)).

For completness I give also the result if $\alpha_{eff}(k)$ satisfies the
RG equation :

$$
\frac{d\alpha_{eff}}{d\ ln\ k^2} = \frac{-\beta_0\ \alpha_{eff}^2}{1 -
\frac{\beta_1}{\beta_0} \alpha_{eff}}
$$
where the {\sl inverse} beta function has only two terms. The appropriate
change of variable in this case turns out
to be precisely the one suggested in ref.[3] :

$$
\frac{z}{z_n} = 1 - \frac{\alpha}{\alpha_{eff}(k)}
$$
(with $\alpha \equiv \alpha_{eff}(k = Q))$, which yields for
$\beta_1/\beta_0 > 0$
(using also the convolution theorem) :

$$
R(\alpha) = \frac{\delta}{1 + \delta}\ \alpha + \frac{1}{1 + \delta}
\int_0^{\infty}
dz\ exp\ \left(-\frac{z}{\alpha}\right) \frac{1}{\left(1 -
\frac{z}{z_n}\right)^{1+\delta}}
$$
whereas for $\beta_1/\beta_0 < 0$, where $\alpha_{eff}$ has an {\sl
infinite} IR fixed point,
one gets :

$$
R(\alpha) = \frac{\delta}{1+\delta}\ \alpha + \frac{1}{1+\delta}
\int_0^{z_n} dz\ exp\ \left(-\frac{z}{\alpha}\right)
\frac{1}{\left(1 - \frac{z}{z_n}\right)^{1+\delta}}
$$
and shows that in this case too (with a different $\tilde{C}$) :
$R(\alpha) = R_{PT}(\alpha) - \tilde{C}(-1)^{\delta}
(\Lambda^2/Q^2)^n$.

For an arbitrary $\beta_{eff}$ function with an IR fixed point one
expects however the answer to be of the more general form :

\begin{eqnarray}
R(\alpha) = R_{PT}(\alpha) + (\Lambda^2/Q^2)^n (-\tilde{C}(-1)^{\delta} + C)
\end{eqnarray}
where $C$ and $\tilde{C}$ are real, and independent of $Q$, and $\tilde{C}$
is proportionnal to the renormalon residue. It may be worth mentionning
that taking the
$Q^2$ derivative of both sides of eq.(19), and going to Borel space, one
easily derives
a relation between $R(z)$ and the Borel transform of $\beta_{eff}(\alpha) =
d\alpha/dln\ Q^2$
(with $\alpha \equiv \alpha_{eff} (k = Q))$ :

\begin{eqnarray}
R(z) - \frac{\beta_0}{n} z\ R(z) - \frac{1}{n} \int_0^z dy\ b(z - y) y\ R(y) =
1
\end{eqnarray}
where $b(z)$ is the Borel transform of $b(\alpha) \equiv -
\frac{\beta_{eff}(\alpha)}{\alpha^2} - \beta_0$.
Eq.(20) can be used to rederive the previous two results, and allows to
deal with more
complicated examples as well. Note also that $C$ and $\tilde{C}$, being
independent of $Q$,
have dropped from eq.(20), and in fact any $Q$-dependence in these
coefficients would be
inconsistent with the assumption of the perturbative nature of the coupling
(equivalent
here to assuming that $\beta_{eff}(\alpha)$ coincides with its Borel sum).
\vskip 1truecm
\noindent
{\bf Acknowledgements}

I thank G. Marchesini and A. Mueller for useful discussions.

\end{document}